\documentclass[12pt]{iopart}

\usepackage{color}
\usepackage{dcolumn}
\usepackage{bm}
\usepackage{amsfonts,latexsym,graphicx}

\newcommand{\ket}[1]{|#1\rangle}
\providecommand{\openone}{\leavevmode\hbox{\small1\kern-3.8pt\normalsize1}}
\usepackage{iopams}

\begin{document}

\title[Comparison of non-Markovianity criteria in a qubit under random external fields]
{Comparison of non-Markovianity criteria in a qubit system under random external fields}

\author{Maria Mannone, Rosario Lo Franco, and Giuseppe Compagno}

\address{Dipartimento di Fisica, Universit\`aà di Palermo, via Archirafi 36, 90123 Palermo, Italy}
\ead{rosario.lofranco@unipa.it}

\begin{abstract}
We give the map representing the evolution of a qubit under the action of non-dissipative random external fields. From this map we construct the corresponding master equation that in turn allows us to phenomenologically introduce population damping of the qubit system. We then compare, in this system, the time-regions when non-Markovianity is present on the basis of different criteria both for the non-dissipative and dissipative case. We show that the adopted criteria agree both  in the non-dissipative case and in the presence of population damping. 
\end{abstract}

\maketitle

\section{Introduction \label{intro}}
In quantum systems the dynamics of decoherence, and that of quantum correlations, is qualitatively different if the environment is Markovian (without memory) or non-Markovian (with memory) \cite{rivasbook, chruscinski, chruscinski_confronto}. For example, for composite quantum systems independent non-Markovian environments, entanglement may present revivals \cite{bellomo2007,lofrancocatania2012PhysScr,lofrancoreview} or trapping \cite{bellomo2008trapping,bellomo2010PhysScrManiscalco} defending it against sudden death \cite{yu2009Science}. Non-Markovian systems are utilized in several physical contexts such as quantum optics \cite{breuer_petruccione}, solid-state physics \cite{Lai}, quantum chemistry \cite{plenio2} and quantum information processing \cite{aharonov}.
It is therefore essential to establish criteria to identify and quantify the non-Markovian behavior in an open quantum system. Among the criteria, one introduced by Breuer-Laine-Piilo (BLP), is based on the concept of temporary flow of information from the environment back into the system and quantify non-Markovianity as an increase in the distinguishability of two evolving quantum states \cite{breuer2009}. A second one, due to Rivas-Huelga-Plenio (RHP), instead measures the deviation of the dynamical map from divisibility \cite{plenio}. A third one has been also proposed by Andersson-Cresser-Hall (ACH) that uses the negative decoherence rates, appearing in the master equation, as a primary measure to completely characterize non-Markovianity \cite{andersson}. An all-optical experiment has been recently developed to control transitions from Markovian to non-Markovian dynamics \cite{Liu}. 

A natural question is then if the different criteria agree in identifying non-Markovian behaviors in the system dynamics. It has been shown that, for a qubit coupled to environments via the Jaynes-Cummings or dephasing models, the BLP and RHP criteria have exactly the same non-Markovian time-evolution intervals and therefore are equivalent \cite{zeng}. In an analysis performed for a driven qubit in a structured environment it has been suggested that the two measures may disagree \cite{haikka2011PRA} and successively it has been shown both for a classical and a quantum toy-model \cite{chruscinski_confronto}. Comparisons among the three criteria, including the ACH one, showing possible non-equivalence in realistic systems are instead still missing.
 
In this paper we address this issue. In particular our aim is to verify, for a realistic physical system made of a qubit subject to random external fields both with and without dissipation, if the BLP, RHP and ACH criteria give concordant results in individuating the non-Markovian time-regions in the system dynamics.

\section{Model}
We consider a realistic system made of a qubit subject to random external fields both in a non-dissipative and in a dissipative case. In the following we describe the two cases.

\subsection{Non-dissipative random external fields}
Our system is a qubit interacting with an environment constituted by a classical field mode with fixed amplitude but with random phase equal either to zero or to $\pi$ with probability $p=1/2$. This model has been introduced to study the possibility of revivals of quantum correlations in absence of back-action \cite{compagno_andersson} and describes a special case of a qubit subject to a phase noisy laser \cite{andersson_master_equation,bellomo2012PhysScripErika}. The dynamical map is of the random external fields type \cite{alicki,zyczowski} and, in the qubit basis $\{\ket{1},\ket{2}\}$, is written as \cite{compagno_andersson}
\begin{equation}\label{dinamica p=1/2}
\Lambda(t,0)\rho(0)=\frac{1}{2}\sum_{i=1}^2 U_i(t)\rho(0)U_i^\dag(t),
\end{equation}
where
\begin{equation}\label{U}
U_i(t)=\left(\begin{array}{@{}cc@{}}
\cos(\lambda t) & e^{-i\phi_i}\sin(\lambda t) \\ -e^{i\phi_i}\sin(\lambda t) & \cos(\lambda t)
\end{array}
\right),
\end{equation}
with $i=1,2$ and $\phi_1=0$, $\phi_2=\pi$. $U_i(t)=e^{-iH_it/\hbar}$ is the time-evolution operator associated to the Hamiltonian $H_i=i\hbar \lambda(\sigma_+e^{-i\phi_i}-\sigma_-e^{i\phi_i})$, where $\sigma_+$, $\sigma_-$ are the qubit raising and lowering operators and $\lambda$ is the qubit-field coupling constant that depends on the field amplitude. The Hamiltonian $H_i$ is given in the interaction picture (rotating frame) at the qubit-field resonant frequency $\omega$.

In order to use the non-Markovianity measures introduced above, the knowledge of both dynamical map and master equation is required. In our model we directly have the map and we also have to construct the corresponding master equation. To obtain the master equation starting from the map of Eq.~(\ref{dinamica p=1/2}) we follow the procedure proposed in Ref.~\cite{andersson_master_equation} which gives (details of calculations are reported in the Appendix)
\begin{equation}\label{master equation dinamica stocastica con p2=1/2}
\mathrm{d}\rho/\mathrm{d}\tau =L\rho(\tau)=\tan 2\tau(\sigma_y\rho\sigma_y-\rho),
\end{equation}
where $\tau=\lambda t$ is a dimensionless time. It is worth to observe that this form of master equation, associated to our system, presents a time-dependent rate, $\tan(2\tau)$, which is the same that has been previously introduced only formally in a general master equation to study non-Markovian behavior \cite{breuer2009, plenio}.

\subsection{Dissipative case}
The model of random external fields described above is non-dissipative and can be generalized to a dissipative case. Although it is not obvious to introduce a source of dissipation directly into the map, it is simple to do into the master equation. We phenomenologically add population damping with rate $\gamma$, in the standard Lindblad form with generator $\gamma\sigma_-$ \cite{nielsen}, into the master equation of Eq.~(\ref{master equation dinamica stocastica con p2=1/2}) which now becomes
\begin{eqnarray}\label{master equation damping}
\mathrm{d}\rho/\mathrm{d}\tau &=& L\rho(\tau)=\tan2\tau(\sigma_y\rho\sigma_y-\rho)\nonumber \\
&& + \tilde{\gamma}(\sigma_-\rho\sigma_+-\rho\sigma_+\sigma_-/2 -\sigma_+\sigma_-\rho/2),
\end{eqnarray}
where $\tilde{\gamma}=\gamma/\lambda$ is a dimensionless decay rate. In the following, we shall use the map of Eq.~(\ref{dinamica p=1/2}) and the master equations of Eqs.~(\ref{master equation dinamica stocastica con p2=1/2}) and (\ref{master equation damping}) to analyze if the different criteria individuate the same time-regions when non-Markovian behavior occurs.

\section{Comparison among the criteria in the non-dissipative case}
We shall first apply the three non-Markovianity criteria (BLP, RHP and ACH) to the case of non-dissipative random external fields.

\subsection{BLP criterion}
The BLP criterion is based on the distinguishability of two evolving quantum states quantified by the trace distance \cite{breuer2009}, that is
$D(\rho_1(t),\rho_2(t))=\frac{1}{2}\|\rho_1(t)-\rho_2(t)\|_1$ where $\|\hat{A}\|_1\equiv\mathrm{Tr}\sqrt{\hat{A}^\dag \hat{A}}$, $\rho_i(t)=\Lambda(t,0)\rho_i$ ($i=1,2$), whose variation rate is
\begin{equation}\label{sigma}
\sigma(t)=\mathrm{d}D(\rho_1(t),\rho_2(t))/\mathrm{d}t.
\end{equation}
The dynamical map $\Lambda(t,0)$ is non-Markovian, according to BLP, if there exists a pair of initial states $\rho_1$, $\rho_2$ such that for some time $t > 0$ the distinguishability of the two states increases, that is $\sigma(t)>0$. This is interpreted as a flow of information from the environment back to the system which enhances the possibility of distinguishing the two states. 

Let us apply this criterion to the model of non-dissipative random external fields. Chosen two arbitrary initial states
\begin{equation}\label{stato_iniziale_non_diag_1}
\rho_1=
\left(
\begin{array}{@{}cc@{}}
\omega & \alpha e^{i\varphi_1} \\ \alpha e^{-i\varphi_1} & 1-\omega
\end{array}
\right),
\quad
\rho_2=
\left(
\begin{array}{@{}cc@{}}
\mu & \beta e^{i\varphi_2} \\ \beta e^{-i\varphi_2} & 1-\mu
\end{array}
\right),
\end{equation}
and substituting them into Eq.~(\ref{sigma}) we obtain 
\begin{equation}\label{sigma fasi generiche}
\sigma(\tau)=-\sqrt{a}\sin(4\tau)/|b|,
\end{equation}
where $a=(\mu-\omega)^2+(\alpha\cos\varphi_1-\beta\cos\varphi_2)^2$ and $b=\cos^22\tau+\alpha\sin\varphi_1-\beta\sin\varphi_2$. The sign of this quantity does not depend on the value of the parameters of the initial states and thus permits a general comparison with the other criteria. In particular it is readily found that $\sigma(\tau)>0$ (i.e., the dynamics exhibits non-Markovianity) when $\pi/4+k(\pi/2)<\tau<(k+1)\pi/2$, where $k$ is a non-negative integer number.

\subsection{RHP criterion}
The RHP criterion is based on the divisibility of a dynamical map and it is independent of the system state. If the map $\Lambda(t,0)$ is divisible, it satisfies the condition $\Lambda_{(t+\epsilon,0)}=\Lambda_{(t+\epsilon,t)}\Lambda_{(t,0)}$ ($\epsilon$ is a time interval) that is usually attributed to Markovian evolution. It is possible to show that the map $\Lambda(t,0)$ is completely positive (CP), and then divisible, if and only if $(\Lambda_{t+\epsilon,t}\otimes\openone_2)|\Phi\rangle\langle\Phi|\geq0$, where $\ket{\Phi}$ is a maximally entangled state of two qubits (one is subject to the map while the other is the isolated ancilla) and $\openone_2$ is the two-dimensional identity matrix \cite{plenio}. For a qubit subject to a master equation $\mathrm{d}\rho/\mathrm{d}t=L_t(\rho)$, where $L_t$ is a Lindblad operator, in the limit of $\epsilon\rightarrow 0$ the solution (dynamical map) of this equation formally tends to  
$\Lambda_{t+\epsilon,t}\rightarrow e^{L_t\epsilon}$. Expanding this solution up to the first order in $\epsilon$ it is possible to introduce the quantity \cite{plenio}
\begin{equation}\label{plenio g}
g(t)=\lim_{\epsilon\rightarrow 0^+}\frac{\|\left[\openone_4+\epsilon(L\otimes\openone_2)\right]|\Phi\rangle\langle\Phi|\|_1-1}{\epsilon},
\end{equation}
where $\|A\|_1$ indicates the trace norm. It is shown that $g(t)>0$ if and only if the original map $\Lambda(t,0)$ is indivisible, that is exhibits non-Markovian behavior. 

In our case of non-dissipative random external fields, identifying $L_t$ with that of the master equation of Eq.~(\ref{master equation dinamica stocastica con p2=1/2}), we obtain $g(\tau)=-2\tan2\tau$ if $\tan2\tau$ $<0$  and $g(\tau)=0$ otherwise. It is immediately seen that non-Markovian behavior ($g(\tau)>0$) occurs just in the same temporal regions individuated above by the BLP criterion, that is $\pi/4+k(\pi/2)<\tau<(k+1)\pi/2$.

\subsection{ACH criterion}
This criterion is based on the property of complete positivity (divisibility) of the dynamical map deduced through the sign of time-dependent decoherence rates that may appear in the master equation. This criterion is also independent of the system state. Consider a qubit governed by a master equation in the canonical (Lindblad-type) form, in the interaction picture \cite{andersson}
\begin{eqnarray}\label{canonical form}
\frac{d\rho}{d\tau}=\sum_k\gamma_k(\tau)[L_k(\tau)\rho L_k^\dag(\tau)
-\frac{1}{2}L_k^\dag(\tau)L_k(\tau)\rho-\frac{1}{2}\rho L_k^\dag(\tau)L_k(\tau)],
\end{eqnarray}
where the traceless operators $L_k(\tau)$, time-dependent in general, describe different decoherence channels and $\gamma_k(\tau)$ are the corresponding decay rates that can be also time-dependent. The different decay channels are orthogonal in the sense that $\mathrm{Tr}(L_j^\dag L_k)=\delta_{jk}$. If the $\gamma_k(\tau)$ are positive at all times, then the time evolution is completely positive in any time interval with a Markovian behavior. On the other hand, if some of the $\gamma_k(t)$ is negative, the time evolution exhibits non-Markovian behavior that can be then naturally characterized by the function $f_k(\tau)=\min[\gamma_k(\tau),0]$ for each decoherence channel \cite{andersson}. This criterion is conceptually similar to the RHP one and it is convenient due to its immediate application once having the expression of the master equation. 

In the master equation of Eq.~(\ref{master equation dinamica stocastica con p2=1/2}), associated to our model of a qubit under non-dissipative random external fields, the only (dimensionless) decay rate is $\tan2\tau$. Once again we find that the time regions when non-Markovian behavior occurs correspond to the negative values of $\tan2\tau$. 

The above results show agreement  among the three criteria in individuating time-regions of non-Markovianity here considered, in the case of non-dissipative random external fields.

\section{Comparison among the criteria in the dissipative case}
We now analyze RHP and ACH criteria in the case of a qubit subject to random external fields and to population decay, whose master equation is given in Eq.~(\ref{master equation damping}). We do not consider the BLP criterion that requires the knowledge of the qubit evolution and therefore the solutions of the master equation of Eq.~(\ref{master equation damping}): this will be treated elsewhere. 

The function $g(t)$ of Eq.~(\ref{plenio g}) of the RHP criterion now becomes
\begin{equation}\label{function g with decay RHP case}
g(\tau)=-\tilde{\gamma}/2-\tilde{\gamma}_1(\tau)/2+(\sqrt{2}/4)\left[\tilde{g}_+(\tau)+\tilde{g}_-(\tau)\right],
\end{equation}
where $\tilde{g}_\pm(\tau)\equiv\left\{\tilde{\gamma}^2+\left[\tilde{\gamma}+\tilde{\gamma}_1(\tau)\right]
\left[\tilde{\gamma}_1(\tau)\pm\sqrt{\tilde{\gamma}^2+\tilde{\gamma}_1^2(\tau)}\right]\right\}^{1/2}$ and 
$\tilde{\gamma}_1(\tau)\equiv2\tan 2\tau$.

To use the ACH criterium, we put the master equation of Eq.~(\ref{master equation damping}) into the canonical form of Eq.~(\ref{canonical form}) by using the procedure of Ref.~\cite{andersson}; two orthogonal decay channels arise with rates
\begin{equation}\label{canonicalrates}
\tilde{\gamma}_\pm(\tau)=(\tilde{\gamma}+2\tan2\tau\pm\sqrt{\tilde{\gamma}^2+4\tan^22\tau})/2,
\end{equation}
and corresponding operators $L_\pm=\sum_{i=1,2}U^{(\pm)}_i\sigma_i/\sqrt{2}$, where $\sigma_i$ ($i=1,2$) are the usual Pauli matrices and
\begin{eqnarray}\label{canonicaloperators}
U^{(\pm)}_1&=&\frac{i(-2\tan2\tau \pm \sqrt{\tilde{\gamma}^2+4\tan^22\tau})}{\sqrt{\tilde{\gamma}^2+(2\tan2\tau\mp\sqrt{\tilde{\gamma}^2+4\tan^2 2\tau}})^2},\nonumber\\
U^{(\pm)}_2&=&\tilde{\gamma}/\sqrt{\tilde{\gamma}^2+(2\tan2\tau\mp\sqrt{\tilde{\gamma}^2+4\tan^2 2\tau}})^2
\end{eqnarray}
Being $\tilde{\gamma}_-(\tau)\leq\tilde{\gamma}_+(\tau)$ at any time, the non-Markovianity regions according to ACH are characterized only by the function $f_-(\tau)=\min[\tilde{\gamma}_-(\tau),0]$. From Eq.~(\ref{canonicalrates}), the condition $\tilde{\gamma}_-(\tau)<0$ is satisfied when $4\tilde{\gamma}\tan2\tau<0$ (i.e., $\pi/4+k(\pi/2)<\tau<(k+1)\pi/2$). Therefore, the ACH criterion in the dissipative case individuates non-Markovianity in the same time regions of the previous non-dissipative case.

\begin{figure}
\begin{center}
\includegraphics[width=0.45\textwidth]{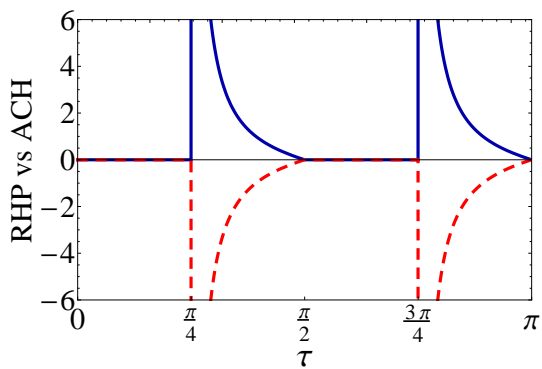}
\end{center}
\caption{\label{Grafico} \footnotesize
Comparison between the function $g(\tau)$ of the RHP criterion (blue solid line) and the function $f(\tau)$ of the ACH criterion (red dashed line) as a function of the dimensionless time $\tau$, for a dimensionless decay rate $\tilde{\gamma}=3$. There is non-Markovianity when $g(\tau)>0$ according to RHP and when $f(\tau)<0$ according to ACH.}
\end{figure}

In this dissipative case, the ACH criterion evidence non-Markovian behavior in the same time regions individuated by the RHP criterion. This is displayed in Fig.~\ref{Grafico}, where it is seen that the function $g(\tau)$ of the RHP criterion is greater than zero exactly when the function $f(\tau)$ of the ACH criterion is lower than zero.
All the above results are independent of the initial state of the system.

\section{Conclusions\label{Conclusions}}
In this paper we have analysed three different criteria (BLP, RHP and ACH) identifying non-Markovian behaviors in a realistic system made of a qubit subject to random external fields, both in a non-dissipative and in a dissipative evolution. We have first exactly obtained the master equation corresponding to the qubit dynamical map of random external fields. We point out that the form of the master equation, associated to our system, contains the time-dependent rate $\tan(2\tau)$ that has been previously inserted only formally in a general master equation to study non-Markovian behavior \cite{breuer2009, plenio}. We have then phenomenologically introduced population damping directly in the master equation associated to the map of random external fields.

We have found, in the non-dissipative case, that the three criteria agree into individuating non-Markovianity time-regions. For the model of random external fields with population decay, both the RHP and ACH criteria individuate the same time regions of non-Markovian behavior.

The results of this paper may provide new insight to the topic of characterizing the non-Markovianity in a realistic open quantum system.

\section*{Acknowledgments}
The authors acknowledge \'{A}ngel Rivas Vargas for fruitful comments and suggestions.

\appendix
\section{Master equation associated to the model of non-dissipative random external fields}
In this Appendix we summarize the steps to obtain the master equation of Eq.~(\ref{master equation dinamica stocastica con p2=1/2}) from the map of random external fields of Eq.~(\ref{dinamica p=1/2}) by following the general procedure described in Ref.~\cite{andersson_master_equation}. 

The general steps are as follows. Let us apply a map $\Lambda(t,0)$ to the basis operators $G_i=\sigma_i/\sqrt{2}$ ($i=0,\ldots,3$), where $\sigma_0=\openone$ and the remaining $\sigma_i$ are the Pauli matrices, and define a matrix $F$ with elements
$F_{kl}\equiv \mathrm{Tr}[G_k\Lambda(t,0)(G_l)]$.
The idea is of constructing a matrix $\dot FF^{-1}$ (or, more generally, $\dot F\tilde{F}$ if $F$ is not invertible). In our case $F$ is invertible and it is possible to calculate the matrix $R$, with elements defined by
\begin{equation}\label{R elements}
R_{ab}=\sum_{rs}(\dot F F^{-1})_{rs}tr[G_r\tau_a^\dag G_s\tau_b],
\end{equation}
where $\tau_a=|\alpha_1\rangle\langle\alpha_2|$, $\tau_b=|\beta_1\rangle\langle\beta_2|$, with $|\alpha_1\rangle$, $|\alpha_2\rangle$ and $|\beta_1\rangle$, $|\beta_2\rangle$ being the qubit basis states $|1\rangle$, $|2\rangle$. The general expression of the master equation is then 
\begin{equation}\label{master equation con matrice R}
L(\rho(\tau))=\dot\rho(\tau)\equiv\sum_{ab}R_{ab}(t)\tau_a\rho(t)\tau_b^\dag,
\end{equation}
where the operators $\tau$ are $\tau_0=|2\rangle\langle2|=\sigma_+\sigma_-$, $\tau_1=|1\rangle\langle1|=\sigma_-\sigma_+$, 
$\tau_2=|2\rangle\langle1|=\sigma_+$ and $\tau_3=|1\rangle\langle2|=\sigma_-$, with $\sigma_\pm=(\sigma_1\pm i\sigma_2)/2$.
In our case of random external fields with the map given in Eq.~(\ref{dinamica p=1/2}), we obtain the matrix $F$
\begin{equation}\label{F}
F=\left( \begin{array}{cccc}
1  & 0 & 0 & 0 \\
0 & \cos 2\tau & 0 & 0 \\
0 & 0 & 1 & 0 \\
0 & 0 & 0 & \cos2\tau
\end{array} \right),
\end{equation}
from which one easily obtains the matrices $F^{-1}$, $\dot F$ and therefore the matrix $\dot F F^{-1}$.
Chosen the basis $\left\{| 2  \rangle\langle  2|,\,| 1  \rangle\langle 1 |,\,|  2 \rangle\langle 1  |,\,| 1  \rangle\langle 2 |\right\}$ and using Eq.~(\ref{R elements}), we find the $R$ matrix as
\begin{equation}
R=\left(
\begin{array}{cccc}
-\tan2\tau & -\tan2\tau & 0 & 0
\\ -\tan2\tau & -\tan2\tau & 0 & 0
\\ 0 & 0 & \tan2\tau & -\tan2\tau
\\ 0 & 0 & -\tan2\tau & \tan2\tau
\end{array}
\right).
\end{equation}
Finally, using Eq.~(\ref{master equation con matrice R}) we obtain the desired master equation
\begin{equation}
\mathrm{d}\rho/\mathrm{d}\tau=\tan2\tau(\sigma_y\rho\sigma_y-\rho).
\end{equation}

\section*{References}
\providecommand{\newblock}{}

\end{document}